\documentclass{osa-article}
\usepackage{physics}
\journal{oe}
\begin{document}

\title{Detection and demultiplexing of cylindrical vector beams enabled by rotational Doppler effect}

\author{Xiaoru Zhang, Junliang Jia,\authormark{*} Kaiyi Zhai, Zehong Chang, Zhenyu Guo and Pei Zhang\authormark{**} %Xuanyi Zhang\authormark{1}
}

\address{MOE Key Laboratory for Nonequilibrium Synthesis and Modulation of Condensed Matter, School of Physics, Xi’an Jiaotong University, Xi’an 710049, China\\
%\authormark{2}Publications Department, The Optical Society (OSA), 2010 Massachusetts Avenue NW, Washington, DC 20036, USA\\
%\authormark{3}Currently with the Department of Electronic Journals, The Optical Society (OSA), 2010 Massachusetts Avenue NW, Washington, DC 20036, USA
}

\email{\authormark{*}jiajl0001@gmail.com\\
\authormark{**}zhangpei@mail.ustc.edu.cn}%% email address is required



\begin{abstract*}
Cylindrical vector beams (CVBs) detection is of vital significance in kinds of studies such as particle observation, mode-division multiplexing. Here we realize a comprehensive detection of cylindrical vector beams based on rotational Doppler effect including analysis of topological charges, amplitudes and phases for mode bases. We construct a mode demultiplexing scheme to obtain the amplitudes, phases in beating signal of collected scattering light by Fourier transformation.The method resolves both absolute values and signs of topological charges of CVB simultaneously, which can not be simply realized by existed polarization examination techniques. It may be of big potential for related researches since an efficient, quantitative and complete scheme to detect CVBs is verified starting from this work.
\end{abstract*}

%%%%%%%%%%%%%%%%%%%%%%%%%%  body  %%%%%%%%%%%%%%%%%%%%%%%%%%
\section{Introduction}
The polarization state of light relates to spin angular momentum, while the helical wavefront of light relates to orbital angular momentum (OAM) \cite{Allen:92}. They both contribute to a higher dimensional beam, cylindrical vector beam (CVB), which is a vectorial solution to the Helmholtz equation for the electric field\cite{Zhan:09}, and has more complicated properties than the homogeneous polarized vortex field. The circular polarized state and spiral phase are classically entangled in CVB, which is reflected by its nonuniform polarization\cite{McLaren:15}. 
Due to its polar symmetrical polarization distribution, CVB has arisen many applications in variety fields such as super-resolution imaging\cite{Chen:13,Segawa:14,Torok:04} and optical-trapping\cite{Kozawa:10,Skelton:13,Donato:12}. 
Besides, the orthogonality between different CVBs also arises amounts of application such as the mode-division multiplexing (MDM) which is widely used in high-capacity communication\cite{Sim:09,Milione:15-2,Wang:16}. 
This technique uses multiple orthogonal modes of the optical field as information carriers. As eigenstates of the polarized optical field with OAM, CVBs act as a carrier for MDM and has a larger capacity compared with the scalar orthogonal sets.

How to effectively generate and detect CVB has become an interesting topic. The simplest way is to fabricate a patterned liquid crystal device called q-plate\cite{Marrecci:06, Cardano:12}. These devices can be set in cavity to make a resonator which directly outputs laser with optical vortex\cite{Wei:19,Naidoo:15}. 
There are also many robust external methods employ interferometers to generate tunable vector beams\cite{liu2018highly}. 
More recently, a new passive generation technique using polarization-selective Gouy phase shifter has been proposed to substitute these interferometers, where it allows simultaneous generation of multiple arbitrary CVBs \cite{Jia:21}. In contrast to the generation, there are relatively fewer researchers concentrating on the detection of CVBs. 
The topological charge (TC) of CVB can be induced by mutual interference pattern between the two orthogonal polarization components of the CVB\cite{Liu:18}. In addition, interferometric mode sorter employing Mach-Zender interferometer makes a connection between the TC of CVB and spatial position, which allows us to obtain the OAM with left or right circular polarizations \cite{Ndagano:17}. The mapping can be further extended to two dimensions with an additional axis representing different polarization states. Simultaneous measurement of OAM and polarization states of CVBs thus is completed through the different points on the plane\cite{Moreno:17}. However, existing detection schemes are erither unable to evaluate the polarization state or their measurement result is not quite quantitative. 
Mode decomposition on the other hand focus on the linear combination of different CVB modes. 
The superposition coefficients can be given by an inner product measurement with a Fourier lens\cite{Ndagano:15}. One alternative way is mode sorter using q-plate or cased interferometer which direct different vector mode into different path\cite{Milione:15,Jia:19}.

Rotational Doppler effect (RDE) is of big significance when involving OAM light. 
When an OAM light illuminates a spinning object with angular velocity $\omega$, the reflected light will carry a frequency shift proportionates to the variation of topological charge \cite{Lavery:13,Fang:17}. RDE can be observed in both monochromatic light and white-light\cite{Lavery:14} and enables plentiful application in the detection of spinning objects\cite{Zhao:16,Guo:21,Deng:19,Zhai:19}. RDE also possesses potential in analyzing OAM spectrum\cite{Zhou:17}. The OAM spectrum analyzer can deduct a full parametric description based on an optical beating signal. Similarly, if we adapt such an analyzer in CVB detection, a full-parameters detection scheme may be constructed simply by a group of time-varying intensities. 

In this paper, we design and demonstrate a scheme to detect CVB with full parameters employing RDE. We first display the RDE induced beating effect in single mode CVB. The beating frequency is proportional to the TC of incident CVB while the visibility and initial phase of intensity variation are regulated by the phase parameter of CVB. We detailedly show that how to distinguish two CVBs with opposite TCs, which is usually degenerated in the analysis of beating signal. The discussion is then expanded to include the multi-mode case, in which the cross term between multiple OAM modes contributes to an extra beating signal in the total intensity. We specially design a spinning object which allows a complete analysis of the one-to-one mapping between the OAM mode TCs and the frequency peaks in the beating signal. Furthermore, the demultiplexing of two degenerate CVBs is achieved.  We perform experiments to verify the complete theoretical scheme. This work enables quantitative, complete, and simultaneous detection of CVB which will show its advantage in CVB mode demultiplexing, particle observing, and other applications.

\section{Single mode detection of vectorial RDE}
Previous research has established a frame where TC of OAM is shifted by light-scattering spinning object and the output frequency is determined by TC shift of incident light\cite{Zhou:16}. The spinning object can be considered as a pure phase modulation $O(\mathrm{r},\phi)$ which corresponds to the roughness of the surface. We expand the modulation function into Fourier series 
\begin{align}
	O(\mathrm{r}, \phi)=\sum_n A_n(\mathrm{r},\phi) \mathrm{e}^{in(\phi-\Omega t)}=\sum_n A_n(\mathrm{r},\phi) \mathrm{e}^{in\phi}\mathrm{e}^{-in\Omega t}\label{eq:1}
\end{align}
where $\Omega$ is the angular velocity of the spinning object and $A_n(\mathrm{r,\phi})$ is the normalized complex amplitude. If a light beam with orbital angular momentum $\ell \hbar$ per photon illuminates the object, the reflected light will be transformed into a superposition of different OAM modes ($\ell^\prime=\ell+n$, where $\ell^\prime$ is the TC of reflected light) and each mode components have a distinct frequency shift $\Delta f=(\ell^\prime-\ell)\Omega/2\pi=n\Omega/2\pi$, which is related only to the object. 
\begin{figure}[h]
	\centering
	\includegraphics[width=\linewidth]{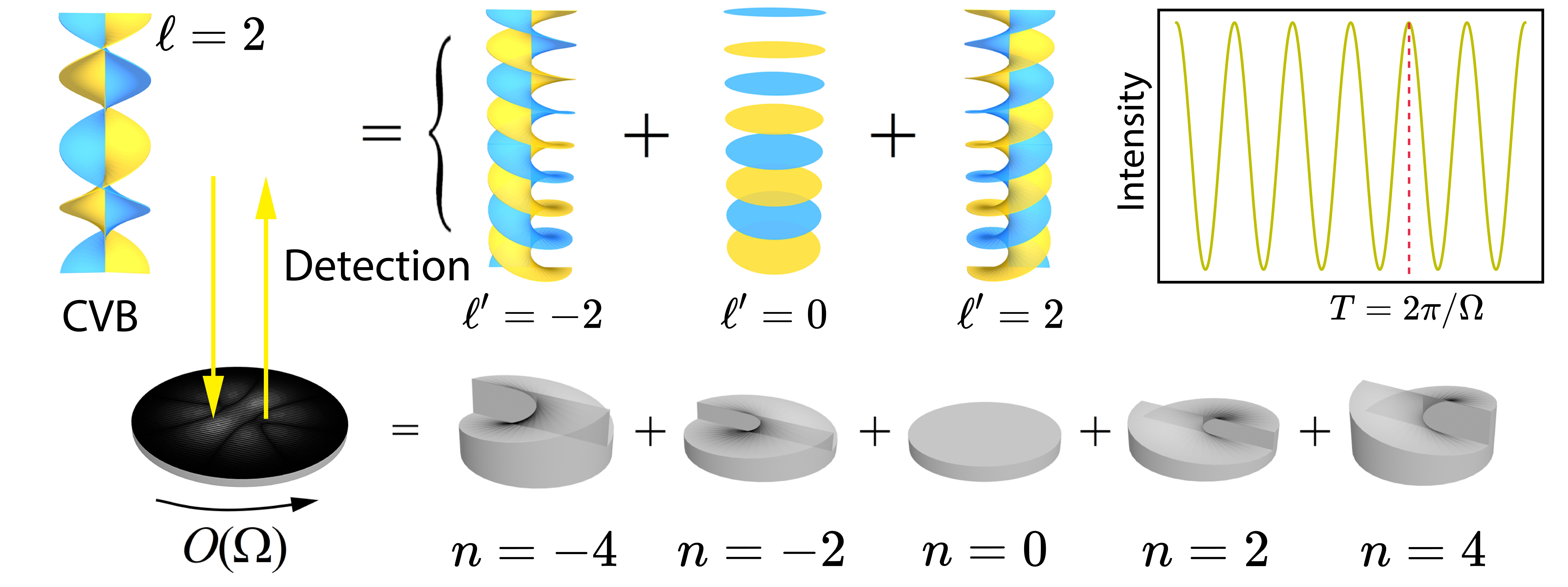}
	\caption{Schematic of detection. A vector beam illuminates a rotating object then induces a beating effect. The TC of the incident CVB is $\ell=2$ while the rotating object can be regarded as a multi-mode spiral phase plate with mode number $n=-4,-2,0,2,4$. The beating effect occurs only on the same OAM basis of the scattered light due to the orthogonality between different OAM modes. The intensity plot in the box demonstrates such beating effect. In this example, only the mode component of the scattered light with $\ell^\prime=-2,0,2$ can produce a variable intensity with beating frequency $f=4\cdot\Omega/2\pi$. 
	}
	\label{RDE demonstrate}
\end{figure}

This article mainly refers to vectorial RDE where CVB undertakes the incident light.
Using the concept of Higher-Order Poincaré Sphere (HOPS)\cite{Milione:11, Milione:12}, the state of arbitrary CVB can be written with
\begin{align}
	\Psi_l(\mathbf{r})=\cos{\upsilon}\mathrm{e}^{-i\gamma}u_\ell(\mathbf{r})\hat{\mathrm{e}}_{\mathbf{R}}+\sin{\upsilon}\mathrm{e}^{i\gamma}u_{-\ell}(\mathbf{r})\hat{\mathrm{e}}_{\mathbf{L}},\label{eq:3}
\end{align}
in which $u_\ell(\mathbf{r})=B_{|\ell|}(r,z)\mathrm{e}^{-i\ell\phi}$ is the complex amplitude of an OAM mode with $\text{TC}=\ell$ and $B_{|\ell|}(r,z)$ is the part of the amplitude that is independent of azimuthal angle $\phi$. $\hat{\mathrm{e}}_{\mathbf{R}}$ and $\hat{\mathrm{e}}_{\mathbf{L}}$ are the unitary vectors correspond to right and left circular polarization. $(2\upsilon,2\gamma)$ are the polar angle and the equator angle of HOPS respectively to characterize the state of polarization. 
As shown in Fig.\ref{RDE demonstrate}, if a CVB illuminates on the spinning object, the vectorial component of the scattered light consists of different OAM modes, and each OAM mode contains two different frequency shifts. Such frequency differences induced a beating effect in the overall intensity signal. Notice also that different modes are orthogonal to each other so that the beating effect only occurs within the same OAM basis. Considering only the horizontal components, the collected intensity of the fundamental mode of scattered light (TC $=0$) is 
\begin{align}
	I_H=\frac{1}{2}\left|B_{|\ell|}\right|^2&\left(\left|A_\ell\right|^2\cos^2{\upsilon}+\left|A_{-\ell}\right|^2\sin^2{\nu}\right.\notag\\
	&+\left|A_\ell A_{-\ell}^*\right|\sin{2\upsilon}\cos(2\ell\Omega t+2\gamma-\xi)\Bigr)\label{eq:7}
\end{align}
where $\xi=\arg(A_\ell A^*_{-\ell})$. Thus, the angular frequency of beating within TC $=0$ basis is $2\ell\Omega t$ which is coincident with twice of TC of CVB. From Eq.\eqref{eq:7}, we see that the light intensity is time-dependent and varies periodically. The beating frequency is now proportional to the TC of the incident CVB while the outcome of frequency shift is only associated with the phase pattern of the object (as shown in Eq.\eqref{eq:1} ). Thus, once we know the angular velocity $\Omega$ of the spinning object, the TC of the incident CVB can be computed throughout the beating frequency. 
%Fig.\ref{RDE demonstrate} shows schematically the process of VRDE and the corresponding beating effect.

Notice that the time-variant term in Eq.\eqref{eq:7} is modified by factor $\cos(2\ell\Omega t+2\gamma-\xi)$ which ranges in $[0,1]$. We can define the visibility $U$ here which describes the magnitude of fluctuation over time of light intensity, the maximum and minimum intensity is denoted as $I_{\text{max}}=\text{max}(I_H), I_{\text{min}}=\text{min}(I_H)$
\begin{equation}
	U=\frac{I_{\text{max}}-I_{\text{min}}}{I_{\text{max}}+I_{\text{min}}} =\frac{\left|A_{\ell}A_{-\ell}^*\right|\sin{2\upsilon}}{\left|A_{\ell}\right|^2\cos^2{\upsilon}+\left|A_{-\ell}\right|^2\sin^2{\upsilon}}\label{eq:8}
\end{equation}
When $\upsilon=0$ or $\upsilon=\pi/2$, the visibility becomes zero. It points out that the time variance is less distinct around the poles of HOPS and eventually vanished at poles. Visibility $U$ reflects the degree of spatial variation of the polarization. At poles of HOPS, spatial polarization distribution becomes homogeneous, and the beating effect disappears accordingly. If the harmonic coefficient of the spinning object satisfied $\left|A_\ell\right|=\left|A_{-\ell}\right|$, Eq.\eqref{eq:8} reduces to $U=\sin{2\upsilon}$. Under this particular condition, it is convenient to determine the polar angle $2\upsilon$ by measuring the visibility of the collected intensity.

\begin{figure}[h]
    \centering
    \includegraphics{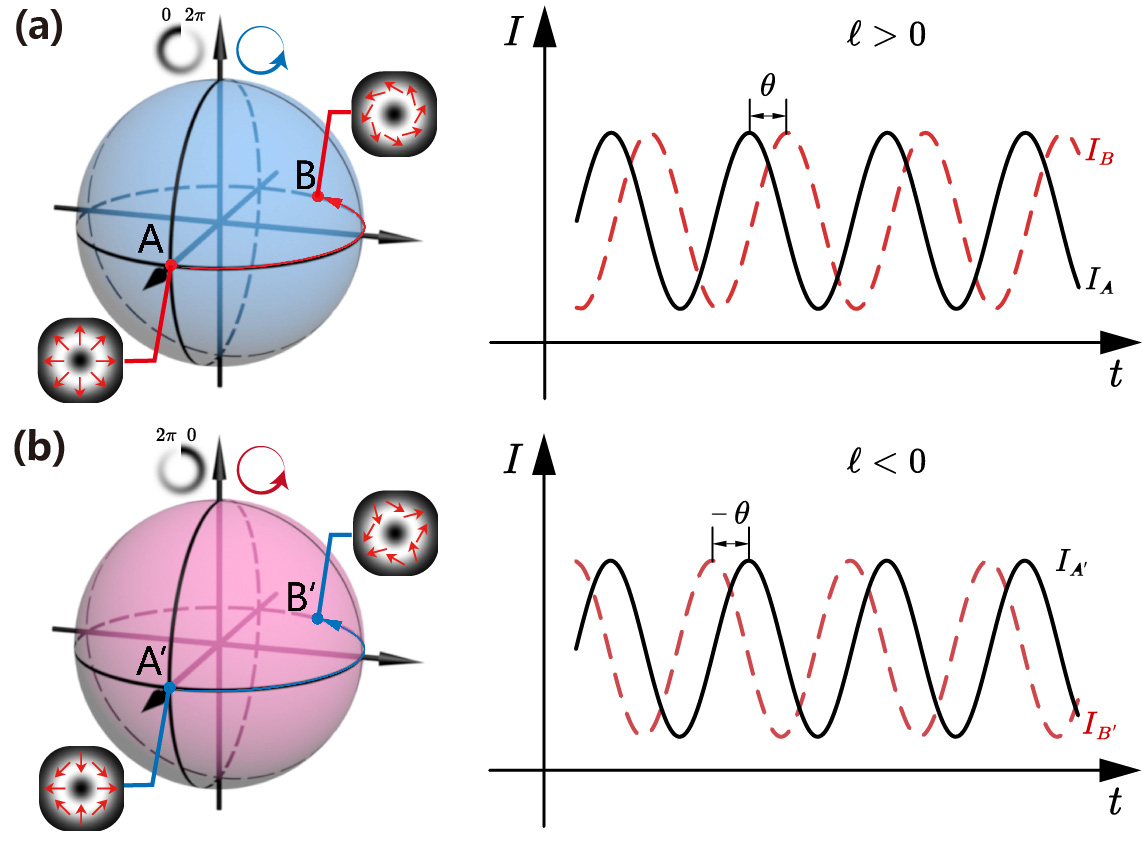}
    \caption{Schematic demonstration of two HOPS entangled with opposite spiral phase. \textbf{(a)} For CVB mode with $\ell=1$, its beating signal experiences positive phase shift under a positive rotation in equator angle. \textbf{(b)} For CVB mode with $\ell=-1$, its beating signal experiences negative phase shift under a positive rotation in equator angle.  }
    \label{fig:degenerate_demo}
\end{figure}

To describe all possible states of CVBs of $|\ell|$, two HOPSs with opposite TC are required\cite{Milione:11}. However, CVBs with the same $|\ell|$ in Eq.\eqref{eq:3} share the same beating frequency and thus each component of the power spectrum is two-fold degenerate. Fig.\ref{fig:degenerate_demo} demonstrates two HOPSs sharing same beating frequency and the difference in polarization distribution between two spheres. Different from common RDE detection, vectorial RDE exhibits different phase shifts in beating after passing two polarizers, which can be utilized to distinguish the rotating direction of the object\cite{Fang:21}.
The phase shift in the intensity signal may also be used to distinguish two frequency-degenerated CVBs. For these two states, the intensity variance over time makes difference only in the sign of the constant phase. Hence, their intensity signal will undergo an opposing phase shift when their equator angle rotates a given degree. 
If the positive rotated angle is denoted by $\theta$, then time variance in the corresponding intensity become 
$ I_H\propto1+\cos{(2\ell\Omega t+2\gamma+\theta-\delta)}\label{eq:10}$. 
It shows that the constant phase now increases linearly as $\theta$ increases if $\ell>0$ and decreases if $\ell<0$. 

\section{Multi-mode demultiplexing of CVB}
In the MDM scheme using vector beam as the information carrier, the transmitted light field is the linear superposition of vector beam with different OAM modes. Here, a multi-mode vector field can be generally expressed as \cite{Rosales_Guzm:18} 
\begin{align}
	E=\sum_\ell C_\ell\left(\cos{\upsilon}\mathrm{e}^{-i\gamma}u_\ell(\mathbf{r})\hat{\mathrm{e}}_R+\sin{\upsilon}\mathrm{e}^{i\gamma}u_{-\ell}(\mathbf{r})\hat{\mathrm{e}}_L\right)
\end{align}
where $C_\ell$ is the superposition coefficient between different modes. 
Similar to before, the rotational phase modulation will still induce different OAM modes in the reflected light and the orthogonality between different modes makes sure the beating happens within the same OAM mode. %Sharing the same notation as before, 
We consider the non-zero coefficients of the scattering object in Eq.\eqref{eq:1}. The intensity of the fundamental mode of reflected light is
\begin{align}
	I_H=\sum_{\ell,\ell^\prime}C_\ell B_{\left|\ell\right|}C_{\ell^\prime}^*B_{|\ell^\prime|}^*&\left[A_\ell A^*_{\ell^\prime}\cos^2{\upsilon}\mathrm{e}^{-i(\ell-\ell^\prime)\Omega t}+A_{-\ell}A_{-\ell^\prime}^*\sin^2{\upsilon}\mathrm{e}^{i(\ell-\ell^\prime)\Omega t}\right. \notag\\
	&\left.+A_{\ell}A_{-\ell^\prime}^*\cos{\nu}\sin{\nu}\mathrm{e}^{-i2\gamma}\mathrm{e}^{-i(\ell+\ell^\prime)\Omega t}+A_{-\ell}A_{\ell^\prime}^*\cos{\upsilon}\sin{\upsilon}\mathrm{e}^{i2\gamma}\mathrm{e}^{i(\ell+\ell^\prime)\Omega t}\right] \label{eq:13}
\end{align}
Obviously, in Eq.\eqref{eq:13}, the overall intensity now contains frequencies that are proportional to cross terms between different TC. 

In order to obtain the mode spectrum of the incident beam, such cross terms need to be excluded. 
For a CVB generated from a Gaussian beam, its radial distribution contains a Gaussian function so that it carries a finite power. 
A spinning object whose Fourier coefficient is proportional to the Laguerre polynomial of the same order can be designed to form radially orthogonal relation with the complex amplitude of the vector beam. The radial integration in Eq.\eqref{eq:13} is thus $ \int^{+\infty}_0B_{\ell}B^*_{\ell^\prime}A_{|\ell|}A^*_{|\ell^\prime|}r\mathrm{d}r\propto\delta_{|\ell||\ell^\prime|}$. The overall intensity now only oscillates at the frequencies that are proportionate to the TC of the incident beam and the beating signals maps with the OAM mode spectrum. However, the frequency degeneration of the single-mode case also occurs in the multi-mode demultiplexing procession. We apply a similar approach to the vertical part of the incident beam. By replacing $\gamma$ with $\gamma+\pi/2$ in Eq.\eqref{eq:13},  we find that the intensity of the vertical components differs from that of the horizontal part with only a sign. Detailed calculation shows that the intensity difference between horizontal and vertical components can be expressed as
\begin{align}
	I_1=\frac{I_H-I_V}{2}=\sum_{\ell=1}^{+\infty}\sin{2\nu}D_\ell\cos{(2\ell\Omega t+\Xi_\ell)}
\end{align}
where $D^{2}_\ell=\left|C_{\ell}\right|^4+\left|C_{-\ell}\right|^4+2\left|C_{\ell}\right|^2\left|C_{-\ell}\right|^2\cos{4\gamma},\tan{\Xi_\ell}=(\left|C_{\ell}\right|^2-\left|C_{-\ell}\right|^2/\left|C_{\ell}\right|^2+\left|C_{-\ell}\right|^2)\tan{2\gamma}$ stands for the amplitude and phase of each mode in the complex spectrum of beating signal. When $\tan{2\gamma}\ne0$, the relative amplitude between CVBs with opposite TC is given by
\begin{align}
	\left|\frac{C_{\ell}}{C_{-\ell}}\right|=\sqrt{\frac{1+\tan{\Xi_\ell}/\tan{2\gamma}}{1-\tan{\Xi_\ell}/\tan{2\gamma}}}\label{eq:deg_amp}
\end{align}

\section{Implementation}
\begin{figure}[h]
	\centering
	\includegraphics[width=\linewidth]{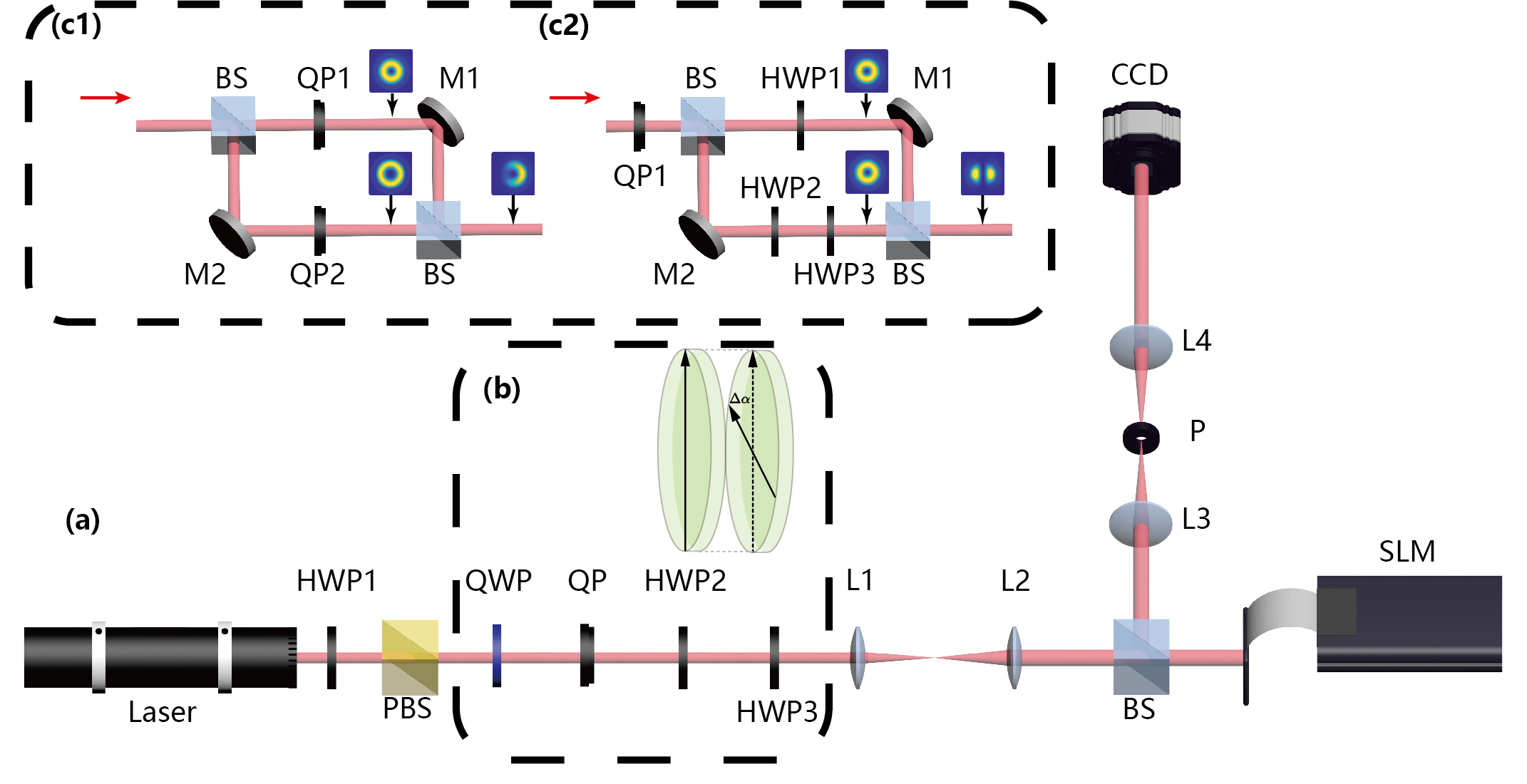}
	\caption{\textbf{(a)} General setup of the experiment. The combination of HWP and PBS is used to regulate the input light intensity. The scattered light from SLM is then directed to a pinhole (P) to select its fundamental mode. The overall intensity is collected through a CCD. \textbf{(b)} Generation setup for single-mode CVB with $\ell=1$. QWP and two HWP are used to adjust the polarization parameter of the CVB. \textbf{(c1)} Generation set up for superposition modes of $\ell=1,2$ with their relative amplitude $C_1:C_2=1:1$. QP1 is a q-plate whose q$=1/2$ and QP2 is a q-plate whose q$=1$. \textbf{(c2)} Generation setup for superposition modes of $\ell=-1,1$. HWP1 and HWP3 are used to adjust the equator angle of the generated CVBs. }
	\label{fig:exp_setup}
\end{figure}

An experimental configuration is implemented to verify the mode detection scheme. Fig.\ref{fig:exp_setup}(a) illustrates the scheme of the experiment. Figure \ref{fig:exp_setup}(b) is the generation setup of single-mode CVB, in which we use a Q-plate with $Q=1/2$ to first generates CVB with $\ell=1$. The polar angle of generated CVB is adjusted by a QWP displayed before the q-plate. Fig.\ref{fig:exp_setup}(c1) and (c2) schematically demonstrated our generation method for multi-mode experiment. We replace (b) with the setup in (c1) and (c2) to superimposed CVBs with different OAM modes. The output beam then illuminates the Spatial Light Modulator (SLM), which is loaded with a computational hologram to simulate the desired spinning object\cite{Zhu:15}. The scattered light then passes through a pinhole to select the fundamental mode and the final light intensity is collected by a charge-coupled device (CCD).

In experiment, we use two HWPs (HWP2, HWP3)%\textbf{(suggest to show b1 and b2 in the position to show results analysis respectively)} 
to change the equator angle $2\gamma$. 
Single HWP transform a point in HOPS at $(2\upsilon, 2\gamma)$ to $(\pi-2\upsilon,-2\gamma+4\alpha)$ where $\alpha$ is the angle between the fast axis of HWP and the horizontal direction. The transformed state passing two HWP is thus $(2\upsilon, 2\gamma+4\Delta \alpha)$ in which $\Delta \alpha=\alpha^\prime-\alpha$ and $\alpha,\alpha^\prime$ is the aligned angle of fast axis of the first and the second HWP as shown in Figure \ref{fig:exp_setup}(b). Two HWP maintain the spiral direction of OAM unchanged.

\begin{figure}[h]
	\centering
	\includegraphics[width=\linewidth]{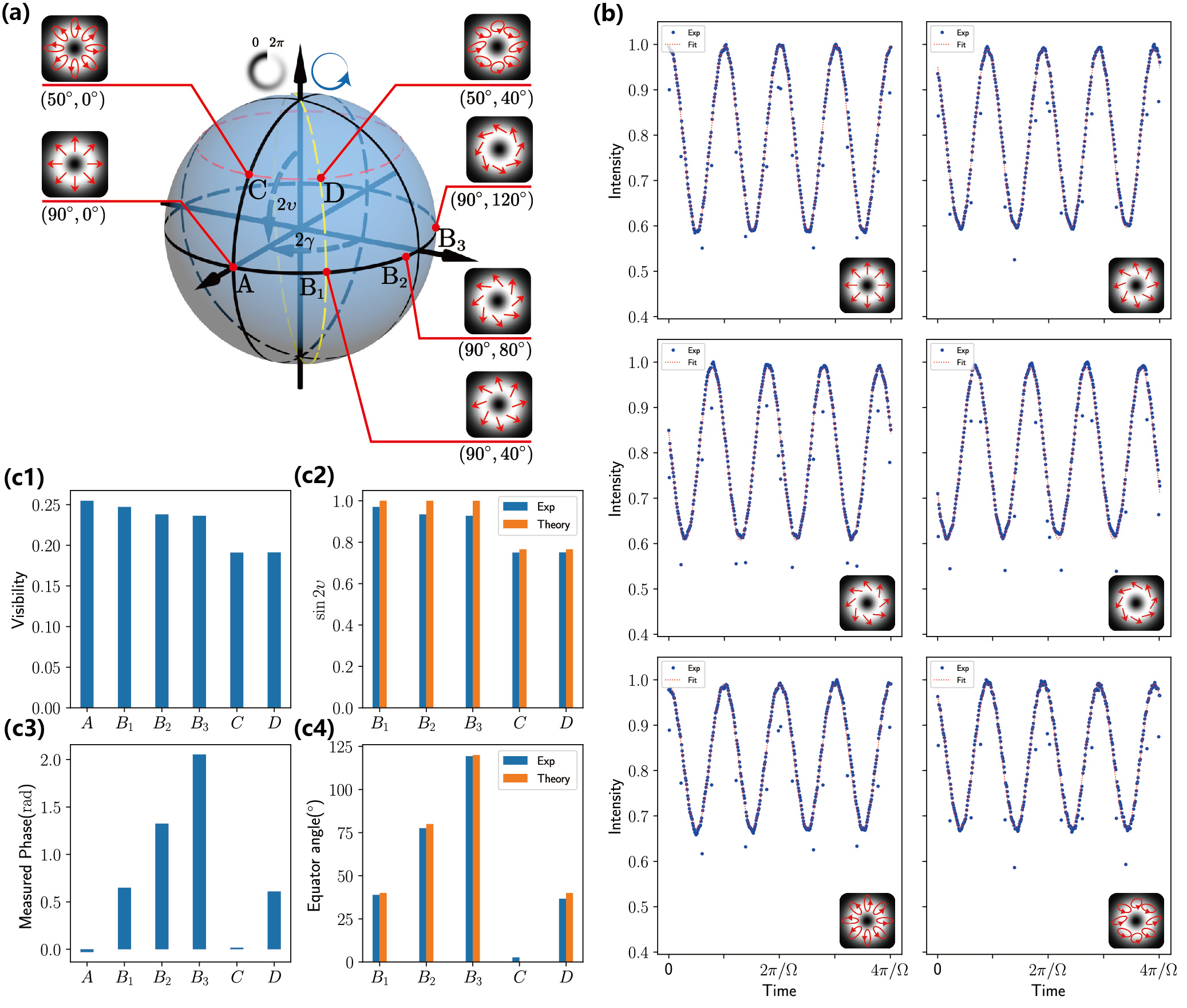}
	\caption{
		\textbf{(a)} Construction of HOPS and some featured points. Point A represents our reference CVB in the succeeding experiment. \textbf{(b)} Measured intensity variation over time for each state. \textbf{(c1)} Visibility of collected intensity. \textbf{(c2)} Experiment value of $\sin{2\upsilon}$ and its comparison with our set up. Here, A is used as the reference light. \textbf{(c3)} Phase of the Fourier transfer of collected intensity. \textbf{(c4)} Experiment value of the equator angle $2\gamma$ and its comparison with theory.
	}
	\label{singlemode_1}
\end{figure}

We measured six featured points on the HOPS of $\ell=1$. Fig.\ref{singlemode_1}(a) displays the polar and equator angle of each point and their polarization distribution. Fig.\ref{singlemode_1}(b) demonstrates the collected intensity of each state. While most of the experimental data fit well with cosine curves, some of them have a large deviation because the q-plate used in our experiment is defective in some positions making the collected intensity sudden decrease at some specific positions. However, the number of affected data is very small and our experimental results continue to be quite valid. Each intensity undergoes two periods when the object rotates one cycle and their initial phase changes with the equator angle, which agrees with our theoretical analysis. The result also verifies that the elliptically polarized field has a smaller range of intensity variation than linear polarization. The smallest value of intensity is not zero implying the existence of time-independent background light. This background light comes from the fact that Fig.\ref{singlemode_1}(b) shows the intensity variance over time for each polarization state. The intensity experiences two full periods for one complete rotation of the object, which consists of our theoretical prediction. Different points experience different phase shifts in their intensities. The measured phase of each intensity plot presented in Fig.\ref{singlemode_1}(c3) shows 
these differences quantitatively. The phase of the intensity signal is acquired by performing a fast Fourier transform (FFT) to the experimental data. Moreover, the extent of intensity variation diminished for the point that is close to the pole. Fig.\ref{singlemode_1}(c1) exhibits the visibility of intensity for each state. Notice that the visibility of radially polarized light is not one as predicted because the modulation of SLM is insufficient and optical devices also introduce additional phase so that its measured phase is also not zero. Therefore, we need to collaborate our system before measurement by performing one measurement first on a reference light with known polarization. The visibility and phase of the reference light serve as a scale. The equator angle of an unknown CVB is given by subtracting the phase of reference light, and the sine value of polar angle is given by the following equation $\sin{2\upsilon_\text{Unk}}=\sin{2\upsilon_{\text{Ref}}}\cdot U_\text{Unk}/U_{\text{Ref}}$. Fig.\ref{singlemode_1}(c2) and (c4) present the final experiment result of polarization parameters. We use the radially polarized field (point A) as the reference light to collaborate the system and attain the polarization parameter of the rest. The result fits well with theoretical expectations, suggesting the feasibility of our method. 

\begin{figure}[h]
	\centering
	\includegraphics[width=0.8\linewidth]{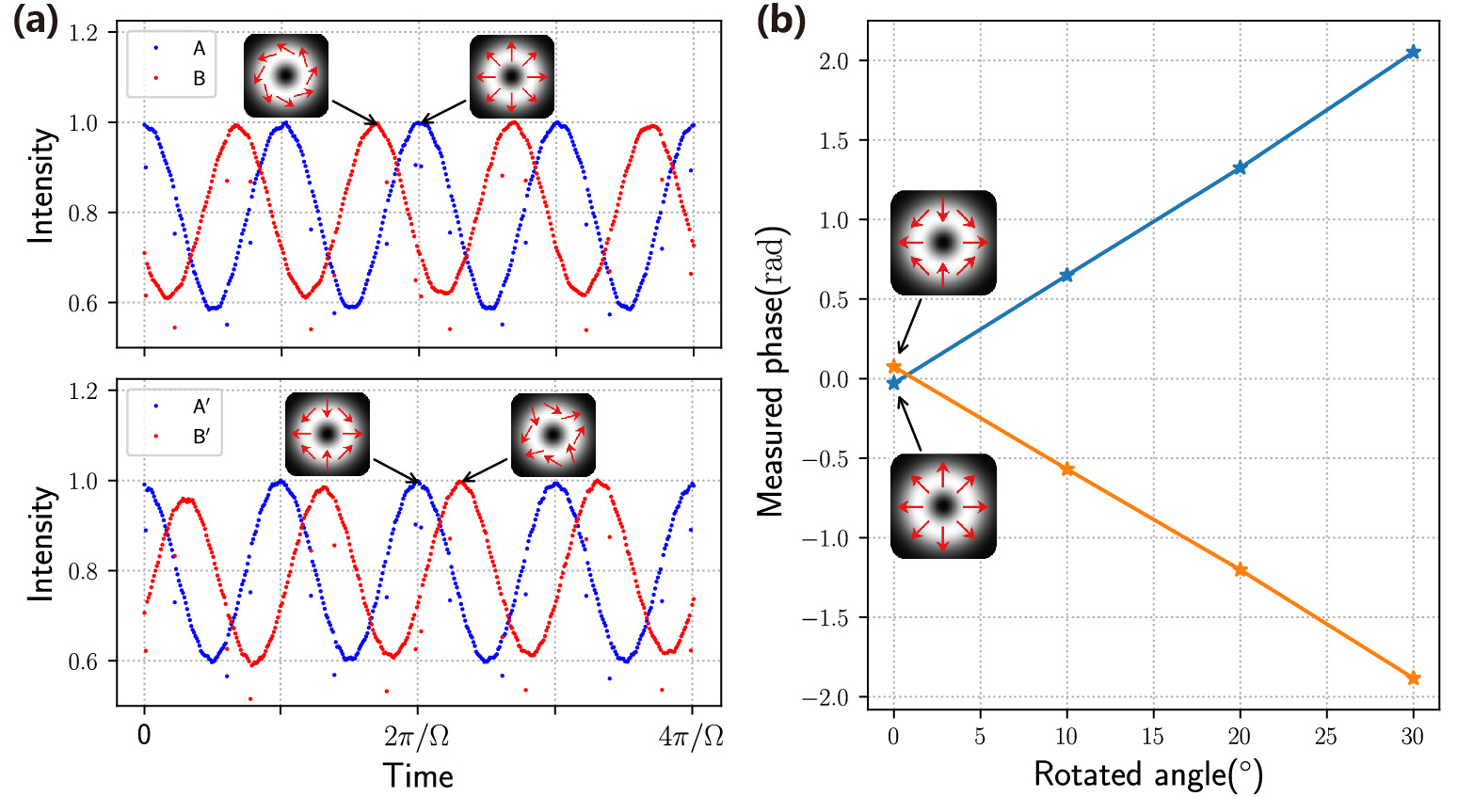}
	\caption{\textbf{(a)} Experiment result of degeneracy in beating effect between radial vector beam and $\pi$-radial vector beam. It also evinces the opposite phase shift in two degenerate spheres when their equator angle change. \textbf{(b)} Relationship between the phase of FFT and rotation angle of HWP, showing two degenerate HOPS experiencing opposite phase shift while rotating the equator angle.}
	\label{singlemode_2}
\end{figure}

We also set up an experiment to verify the previous discussion about the degeneracy between two HOPS. We measured two states on each HOPS with the same polarization parameters presented in Fig.\ref{fig:degenerate_demo}. It is evidenced by Fig.\ref{singlemode_2}(a) that radial (point A) and $\pi$-radial (point A') CVB share identical intensity variation. The experimental result also demonstrates that a rotated state B on the higher-order sphere with $\ell=1$ experiences phase delay while the same rotated state B' on the corresponding sphere with $\ell=-1$ undergoes phase advance. Fig.\ref{singlemode_2}(b) shows the relationship between measured FFT phase and rotation angle of HWP. It is apparent from the result that two degenerate HOPS have opposite phase change with the equator angle.

\begin{figure}[h]
	\centering
	\includegraphics[width=0.8\linewidth]{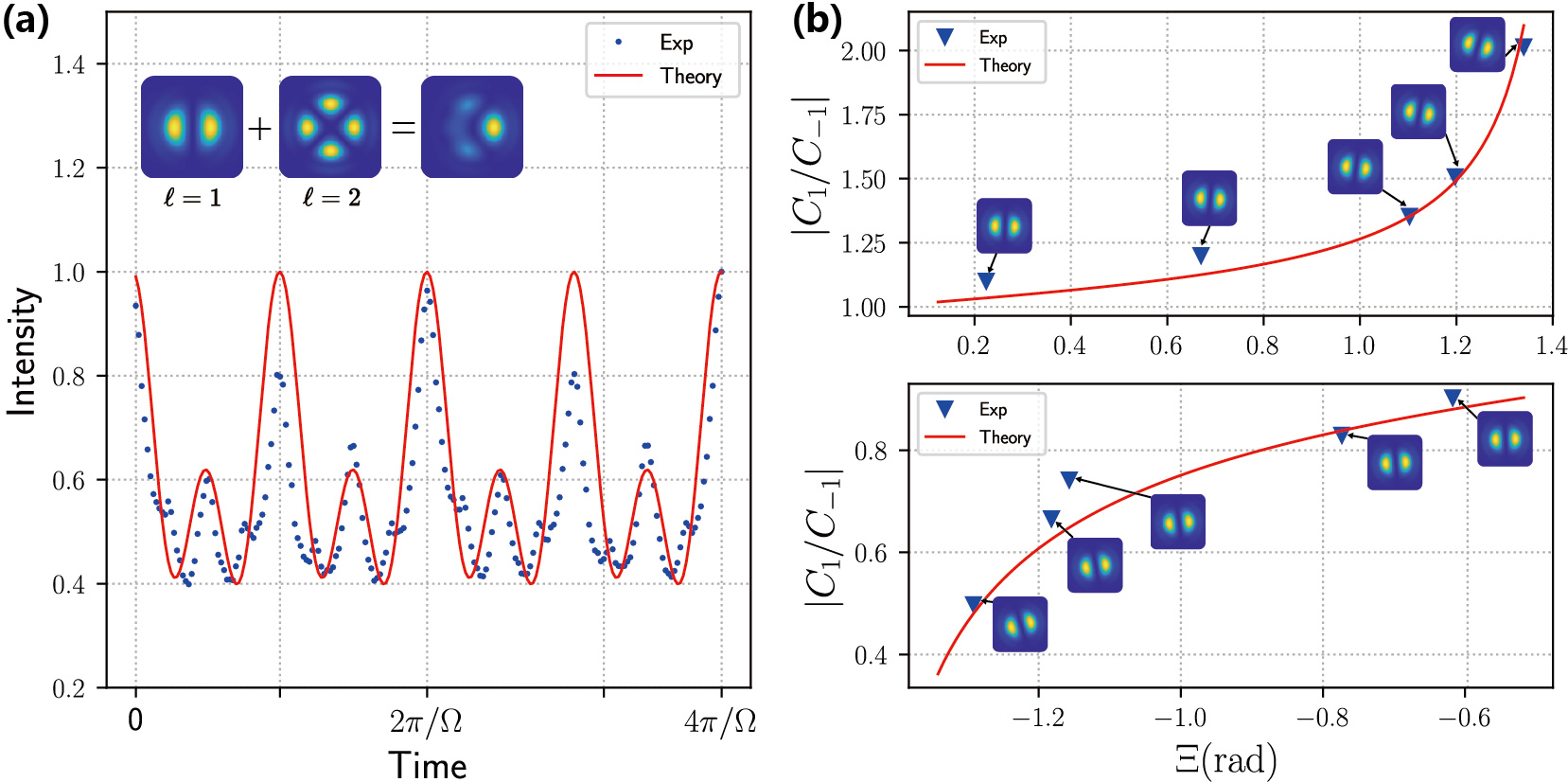}
	\caption{\textbf{(a)} Intensity variance for superimposed states of $\ell=1,2$, whose relative amplitude is $C_1:C_2=1:1$. The intensity distribution of the horizontal components of each superimposed state is also shown in this sub-figure. \textbf{(b)} Demultiplexing of two frequency degenerate modes with TC$=-1,1$. The theoretical prediction of Eq.\eqref{eq:deg_amp} when $\gamma=40^\circ$ is drawn in read line. The intensity distribution for different superposition coefficient is also shown.}
	\label{multimode_exp}
\end{figure}

In deep experiment, we demultiplex the dual superposition states for TC$=1, 2$ and TC$=-1, 1$, respectively, to demonstrate that the proposed method is valid. The experiment set up for generation of these two superposition modes is shown in Fig.\ref{fig:exp_setup}(c1) and (c2). In Fig.\ref{fig:exp_setup}(c1), the horizontally polarized Gaussian beam goes via a beam splitter (BS), then through two q-plates with q$=1/2$ and q$=1$ separately, before being overlaid at the other BS. When a Gaussian beam illuminate a q-plate, the scattered mode with OAM is so called Hypergeometric Gaussian mode\cite{Rubano:19,Karimi:07}. In addition to a Gaussian function, the radial amplitude distribution of a Hypergeometric Gaussian beam is modified by an extra factor of confluent hypergeometric function and radius power term\cite{Kotlyar:08,Kotlyar:05}, and their effect is cancelled by the harmonic coefficients of the spinning object. Vector beams of TC $=1$ and TC $=2$ are superimposed equally in amplitude, and after eliminating the cross term between different mode using the radial orthogonality, the time variation of the light intensity is equally distributed in harmonic terms of $2\ell\Omega$ and $4\ell\Omega$ due to the absence of degenerated modes. Fig.\ref{multimode_exp}(a) presents the intensity variance in theory and our experiment result, both have been normalized. As shown in Fig.\ref{multimode_exp}(a), the experimental data are in good agreement with the theoretical predictions, demonstrating that our approach effectively eliminates the influence of cross terms. In Fig.\ref{fig:exp_setup}(c2), the horizontally polarized light first passes through the q-plate with q$=1/2$, and the generated radially polarized vector beam is subsequently divided into two beams by a BS. 
CVB passes through two half-wave plates with its topological charge remaining unchanged, while the other passes through only one with its topological charge reversed in sign. Two CVBs are then superimposed through another BS. We rotate both HWP3 and HWP1 by $45^\circ$ simultaneously to measurement the vectorial polarization component. 
Fig.\ref{multimode_exp}(b) shows our experimental results for the mode demultiplexing of two degenerate CVBs with their equator angle $\gamma=40^\circ$. The vertical axis represents the relative amplitude between CVB of $\ell=1$ and $\ell=-1$ while the horizontal axis represents the constant phase of intensity variance. We perform an experiment on input light with different relative superimposed amplitude between $(0,2)$ and compare the result measured by an optical power meter with that using Eq.\eqref{eq:deg_amp}. Our method shows good alignment with the actual result. 

\section{Conclusion}
In summary, this study has shown that RDE for a CVB will introduce a beating frequency in the scattered light beam. By introducing a collaboration procedure, the polarization parameters of incident CVB are determined by the visibility and phase of intensity variance. Also, two CVBs with opposite TCs which share the same beating frequency can be distinguished because of opposite phase shifts when their equator angle changes. For RDE of multi-mode CVB, the correspondence between beating signals and mode spectrum is no longer one-to-one due to the presence of cross-beating between different modes. Using a rotating object whose harmonic components form an orthogonal relationship with the radial amplitude distribution of the incident beam, the effect of cross-terms is reduced. The phase of the intensity difference between the horizontal and vertical polarization of the dispersed light is used to demultiplex two degenerate modes.  Analysis of vectorial RDE undertaken here has provided a more quantitative and efficient design in vector beam mode detection and demultiplexing. Based on this paper, a full-parameters detection scheme can be constructed simply by a group of time-varying intensities and reversely can elevate many applications of CVBs.  

%Further research could also be conducted to explore the application of rotational Doppler effect in mode demultiplexing of vector beam.However, One problem lay in the study is the disturbance signals come from the cross term of incident modes. The distribution of cross term in the power spectrum can be diminished by introducing a strong reference light or by modifying the spinning object. The similar frequency degeneracy happens also in multi-mode situation.

%%%%%%%%%%%%%%%%%%%%%% References %%%%%%%%%%%%%%%%%%%%%%%%%

%%%%%%%%%% If using BibTeX:
\bibliography{formal}

\end{document}